\documentclass[letterpaper, 10 pt, conference]{ieeeconf}  

\usepackage{amssymb}
\usepackage[fleqn]{amsmath}
\usepackage{graphicx}

\IEEEoverridecommandlockouts                              

\usepackage{xcolor}
\allowdisplaybreaks

\title{\LARGE \bf
MPC Controller Tuning using Bayesian Optimization Techniques
}

\author{Qiugang Lu$^{1}$, Ranjeet Kumar$^{2}$, and Victor M. Zavala$^{3,*}$
\thanks{*Corresponding author.}
\thanks{$^{1}$Qiugang Lu is with the Department of Chemical Engineering, Texas Tech University, Box 43121, Lubbock, TX 79409, USA
        {\tt\small jay.lu@ttu.edu}}%
\thanks{$^{2}$Ranjeet Kumar is with the Dow Chemical Company, 1254 Enclave Pkwy, Houston, TX 77077, USA
        {\tt\small rkumar27@dow.edu}}%
 \thanks{$^{3}$Victor M. Zavala is with the Department of Chemical and Biological Engineering, University of Wisconsin - Madison, 1415 Engineering Dr, Madison, WI 53706, USA
    	{\tt\small victor.zavala@wisc.edu}}%
}

\begin{document}

\maketitle
\thispagestyle{empty}
\pagestyle{empty}

\begin{abstract}

We present a Bayesian optimization (BO) framework for tuning model predictive controllers (MPC) of central heating, ventilation, and air conditioning (HVAC) plants. This approach treats the functional relationship between the closed-loop performance of MPC and its tuning parameters as a black-box. The approach is motivated by the observation that evaluating the closed-loop performance of MPC by trial-and-error is time-consuming (e.g., every closed-loop simulation can involve solving thousands of optimization problems). The proposed BO framework seeks to quickly identify the optimal tuning parameters by strategically exploring and exploiting the space of the tuning parameters. The effectiveness of the BO framework is demonstrated by using an MPC controller for a central HVAC plant using realistic data. Here, the BO framework tunes back-off terms for thermal storage tanks to minimize year-long closed-loop costs. Simulation results show that BO can find the optimal back-off terms by conducting 13 year-long simulations, which significantly reduces the computational burden of a naive grid search. We also find that the back-off terms obtained with BO reduce the closed-loop costs. 
\end{abstract}

\section{INTRODUCTION}

Model predictive control (MPC) is widely used in industrial systems due to its ability to handle diverse types of constraints, multivariable models, and operational objectives. The performance of MPC depends rather strongly on the controller formulation. Examples of typical tuning parameters that influence performance include the prediction and control horizon, weights in individual states or cost objectives, input rate constraints, and constraint back-off terms \cite{yamashita2016tuning, koller2018stochastic}. Complex and non-intuitive dependencies are typically observed between the tuning parameters of the MPC controller and of its closed-loop  performance;  as such, conducting MPC tuning by trial-and-error or by using heuristics might require a significant number of closed-loop simulations. This represents a problem because a single closed-loop simulation might require the solution of hundreds to thousands of optimization problems. For instance, one is often interested in evaluating the performance of MPC over an entire year of operation or over different operational scenarios.
\\

Self-tuning methods cast the MPC tuning problem as an optimization problem in which the tuning parameters are used to maximize closed-loop performance.  Derivative-free optimization algorithms such as genetic algorithms and particle swarm optimization have been previously proposed to solve the tuning problem. Well-known issues encountered with these techniques include slow progress (thus requiring many simulations) and lack of convergence guarantees. An excellent review of MPC tuning methods can be found in \cite{garriga2010model}. 
\\

Bayesian optimization (BO) is a powerful technique for optimizing computationally-intensive black-box functions \cite{brochu2010a}. BO has been widely used for hyper-parameter tuning of deep learning models \cite{gardner2015scalable}, for design of experiments \cite{greenhill2020bayesian}, and for conducting reinforcement learning tasks \cite{wilson2020using}. BO can also be adapted to accommodate a mixture of continuous and discrete decision variables \cite{brochu2010a} and uses a statistical model to systematically guide exploration and exploitation steps \cite{mockus2012bayesian}.  Exploration aims to evaluate the objective at points in the decision space with the goal of improving the accuracy of a surrogate model of the objective, while exploitation aims to use the surrogate model to identify decisions that reduce (or increase) the objective function.   

In this work, we tackle the MPC tuning problem by using BO techniques. BO approaches have been recently used to tune MPC controllers \cite{forgione2019efficient,lucchini} and for performance-oriented learning of closed-loop dynamical systems \cite{bansal2017goal,bemporad2}. Our work is motivated by an MPC application to heating, ventilation, and air conditioning (HVAC) plants. The cost of HVAC systems is strongly affected by disturbances that cannot be forecast perfectly (demands of electrical power and hot and cold water). Errors in disturbance forecasts result in frequent constraint violations in thermal storage levels (overfilling or dry-up) that ultimately translate in decreased economic performance. Adding a back-off term to the storage levels has been shown to provide a suitable approach to deal with these issues \cite{kumar2020stochastic} and resembles constraint back-off approaches recently explored in the MPC literature  \cite{rafiei2018stochastic}.  Unfortunately, tuning these back-off terms requires extensive simulations. Every closed-loop simulation requires solving over 8,700 optimization problems and is time-consuming (a single simulation requires 2 hours of wall-clock time). Our results indicate that BO can find optimal back-off terms by conducting a total of 13 closed-loop simulations, which significantly reduces the computational burden of naive tuning approaches. We also find that the optimal back-off terms obtained with BO reduce closed-loop HVAC costs.

\section{MPC Tuning using Bayesian Optimization}
\label{Section: II}

We formulate the MPC tuning problem as:
\begin{subequations}\label{eq: MPC_tuning}
\begin{eqnarray}
	&&\min_{\xi}~~f(\xi)  \label{eq: tuning_obj} \\
	&&s.t.~~\xi\in{\Omega}. \label{eq: tuning_cons}
\end{eqnarray}
\end{subequations}
Here, $f(\cdot)=\sum_{i=1}^{n_f}w_if_{i}(\cdot)$ is the tuning objective (assumed to be a smooth function), $f_{i}(\cdot)$ represents the $i$-th tuning objective, $n_f$ is the number of objectives, $\xi \in \mathbb{R}^d$ are the tuning parameters  (e.g., control horizon, weights, back-off terms),  $w_i$ is a user-specified weight capturing relative importance of each tuning objective $f_i(\cdot)$, and $\Omega\subseteq \mathbb{R}^d$ is the space of possible tuning parameters. Examples of tuning objectives of interest  include closed-loop tracking, economic costs, and constraint violations \cite{garriga2010model}. In general, there is no explicit form between the objective function $f(\xi)$ and the parameters $\xi$; as such, the objective function is treated as a black-box function that can only be evaluated via simulation. Extensive simulations might be required to cover the tuning space $\Omega$ in searching for parameters that minimize the tuning objective; as such, we want to derive an algorithm that can more systematically explore the space. 

\begin{figure}
	\centering
	\includegraphics[width=0.8\linewidth]{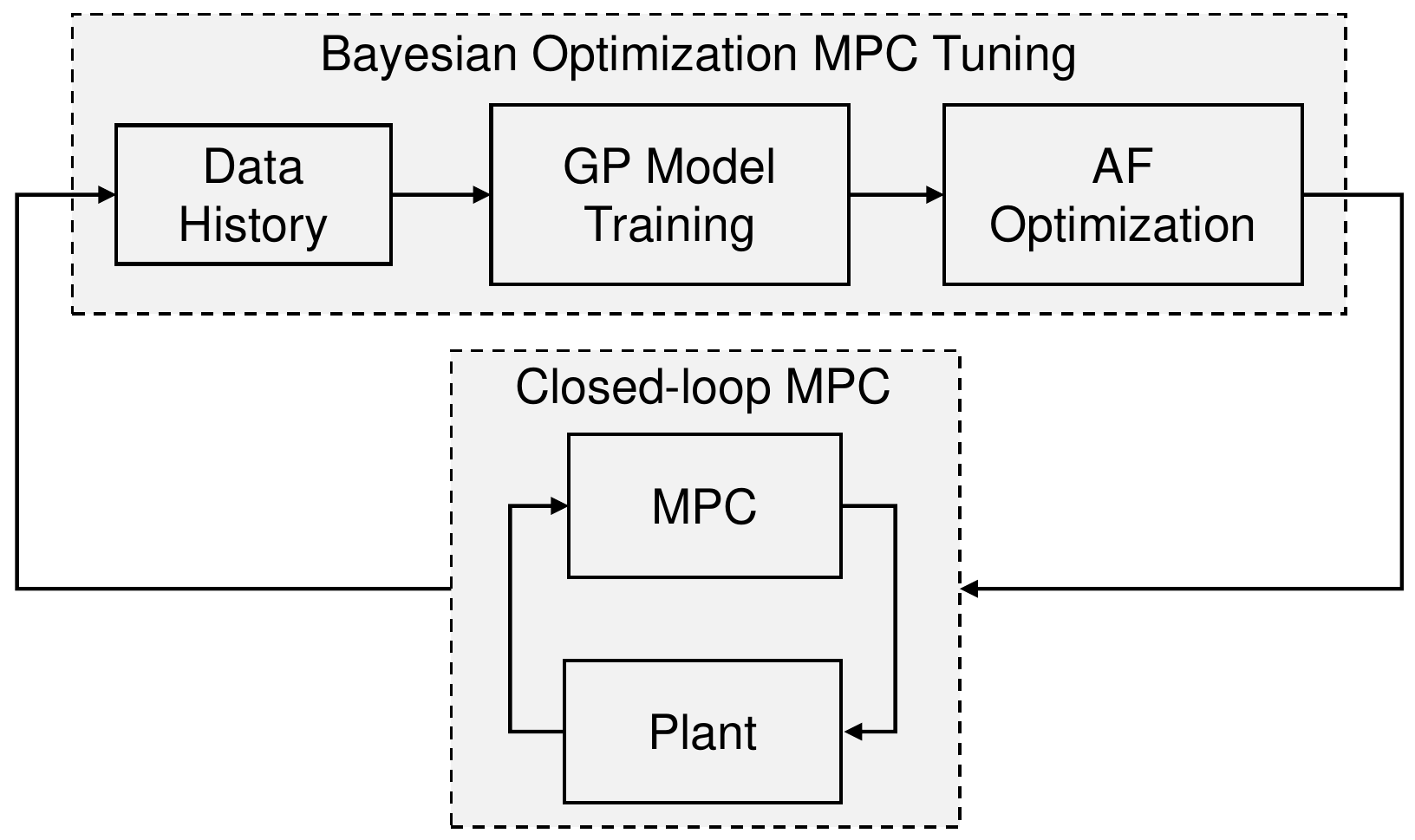}
	\caption{Schematic representation of BO framework for MPC tuning.}
	\label{fig: BO_tuning}
\end{figure}

BO is a family of algorithms for solving black-box optimization problems of the form \eqref{eq: MPC_tuning}.  Given a set of initial $n$ observations of the black-box objective at sample points (denoted as $\{\xi_{1:n},f(\xi_{1:n})\}$), BO constructs a surrogate (statistical) model of the objective function $f(\cdot)$. The statistical model (typically a gaussian process model) provides a posterior distribution of the objective function $f(\cdot)$. The posterior distribution is used to construct an acquisition function (AF); the AF measures the uncertainty (e.g., variance) and performance (e.g., mean) of the objective function at any unexplored point $\xi \in \Omega$. Thus, minimizing the AF provides a natural mechanism to select an optimal sampling point $\xi_{n+1}$. The new observation $\{\xi_{n+1},f(\xi_{n+1})\}$ is then added into the dataset to update the statistical model.  In general, minimizing the AF is a much simpler problem than the original optimization problem.  As the iterations continue and more data is gathered, the surrogate model approaches the true function in the neighborhood of a solution and the subsequent sampling points converge to a true solution. The global convergence and the convergence rate of BO have been thoroughly studied in the literature \cite{wilson2011convergence}. Figure \ref{fig: BO_tuning} provides a schematic representation of the MPC tuning problem solved using BO. 

We use gaussian process (GP) model as the surrogate model to approximate the objective function.  Specifically, we assume that the surrogate function $f(\xi)$ has a GP prior of the form $f(\xi)\sim \mathcal{GP}(m(\xi),k(\xi,\xi^{\prime}))$, where $m(\xi)=0$ is the prior mean and $k(\xi,\xi^{\prime})$ is the covariance function between $\xi$ and $\xi^{\prime}$. We choose the Matern kernel function as the covariance function:
\begin{align}
	&k(\xi,\xi^{\prime})=\nonumber\\
	&\frac{1}{\Gamma(\nu)2^{\nu-1}}\left(\frac{\sqrt{2\nu}}{l}d(\xi,\xi^{\prime})\right)^{\nu}K_{\nu}\left(\frac{\sqrt{2\nu}}{l}d(\xi,\xi^{\prime})\right)
\end{align}
where $\nu$ is the smoothness parameter, $K_\nu (\cdot)$ is the modified Bessel function,  $\Gamma(\cdot)$ is the gamma function, $d(\cdot,\cdot)$ is the Euclidean distance, and $l$ is length-scale parameter. With $n$ samples of data $\{\xi_{1:n},f_{1:n}\}$ where $f_{1:n}=f(\mathbf{\xi}_{1:n})$,  the joint distribution of function values $f_{1:n}$ is a multivariate Gaussian $\mathcal{N}(\mathbf{0},\mathbf{K})$ with zero mean and covariance matrix $\mathbf{K}(\xi_{1:n},\xi_{1:n})\in \mathbb{R}^{n\times n}$, where $\mathbf{K}_{i,j}=k(\xi_i,\xi_j)$. For any other candidate point $\xi$, the corresponding function value $f(\xi)$ and available data samples $f_{1:n}$
are jointly Gaussian: 
\begin{equation}
	\left[
	\begin{array}{c}
     	f_{1:n} \\
		f(\xi)
	\end{array}
	\right]\sim \mathcal{N}\left(
	\mathbf{0}\left[
	\begin{array}{cc}
		\mathbf{K}(\xi_{1:n},\xi_{1:n}) + \sigma^{2}\mathbf{I}& \mathbf{K}(\xi_{1:n},\xi)  \\
		\mathbf{K}(\xi_{1:n},\xi)^{T} & \mathbf{K}(\xi,\xi)
	\end{array}
	\right]
	\right),
\end{equation}
where $\sigma^{2}$ is the noise level of the measurement. We thus have that the posterior $p(f(\xi)|f_{1:n},\xi_{1:n},\xi)$ is Gaussian with mean and covariance \cite{brochu2010a}:
\begin{subequations}
\begin{align}
\mu &= \mathbf{K}(\xi,\xi_{1:n})\left[\mathbf{\mathbf{K}}(\xi_{1:n},\xi_{1:n}) + \sigma^{2}I \right]^{-1}\xi_{1:n}, \\
\Sigma&=\mathbf{K}(\xi,\xi) + \sigma^{2}\mathbf{I}\nonumber \\ & \quad - \mathbf{K}(\xi,\xi_{1:n})^{T}\left[\mathbf{\mathbf{K}}(\xi_{1:n},\xi_{1:n}) + \sigma^{2}I \right]^{-1}\mathbf{K}(\xi_{1:n},\xi).
\end{align}
\end{subequations}
The above posterior distribution provides an explicit representation of the mean and variance for the objective function.  These are used to construct an acquisition function (AF) that is used to direct the search for the optimal $\xi$. In this work, we  use the lower confidence bound (LCB) as our AF:
\begin{eqnarray}
	\text{LCB}(\xi)=\mu(\xi)-\kappa \sigma(\xi), \label{UCB}
\end{eqnarray}
where $\kappa>0$ is a hyper-parameter, $\mu(\xi)$ is the posterior mean of $f(\xi)$ and $\sigma(\xi)$ is the posterior variance of $f(\xi)$. The next sampling point $\xi_{n+1}$ is obtained by solving the problem:
\begin{subequations}
\begin{eqnarray}
	\min&~~\text{LCB}(\xi) \\
	s.t. &~~ \xi\in{\Omega}\subseteq\mathbb{R}^{d},
\end{eqnarray}
\end{subequations}
Note that the AF considers both the predicted mean and variance; when the weight parameter $\kappa$ is large, the step $\xi_{n+1}$ seeks regions of large variance (known as exploration step). On the other hand, with a small weight $\kappa$, $\xi_{n+1}$ seeks regions that reduce the predicted mean (known as exploitation step). After solving the AF minimization problem, one evaluates the objective function at $\xi_{n+1}$ and incorporates the observation $(\xi_{n+1},f_{n+1})$ into the dataset. A new GP model is re-trained based on $\{\xi_{1:n+1},f_{1:n+1}\}$ and this is used to obtain the next sampling point via AF minimization.

\section{Case Study: MPC Tuning for HVAC Plants}
\label{Section: III}

Thermal energy storage (TES) for chilled/hot water is used to shift energy loads of an HVAC plant to off-peak hours in order to reduce electricity costs and to mitigate peak demands \cite{rawlings2018economic}. Energy demands and prices are difficult to forecast and errors often result in violations of TES capacity limits (overflow or drying up of water tanks). A strategy to mitigate these violations consists of using a reserved buffer (by adding a back-off term on the storage constraints). Currently,  these back-off terms are selected by manual search, which requires repeated simulations of the closed-loop system. This approach is time-consuming as it involves year-long simulations. 

In this case study, we leverage the MPC formulation proposed in \cite{kumar2020stochastic} and build a BO framework for tuning TES back-off terms. In the HVAC plant, a chiller subplant produces chilled water and a heat recovery (HR) chiller subplant produces both chilled water and hot water; a hot water generator produces hot water; cooling towers are used to decrease temperature of water purchased from the market; a dump heat exchanger (dump HX) rejects heat from the hot water; and storage tanks (one for chilled water and one for hot water) are used as the TES. 
The MPC controller seeks to determine hourly operating loads for each unit in such a way that the HVAC plant satisfies the demands of chilled and hot water from multiple buildings of a university campus. The objective of the MPC is to minimize the total cost of the utilities (electricity, water, and natural gas) purchased from the market. Electricity is charged based on time-varying prices, while water and natural gas usage are charged at constant prices. 

The HVAC plant cost includes the following items: (i) electricity required for the equipment operation and charged based on hourly time-varying prices, $\pi^{e}_{t}$, (ii) water required to make up for evaporative losses of water in the cooling towers and purchased at a fixed price $\pi^{w}_{t}$ = \$0.009/gal, (iii) natural gas required for the operation of hot water generator to satisfy the campus heating load and purchased at a fixed price of $\pi^{ng}_{t}$ = \$0.018/kWh, and (iv) the peak electrical demand charges for each month charged at a high rate of $\pi^{D}$ = \$4.5/kW. 

\begin{figure}[!ht]
\centering
\includegraphics[width=0.45\textwidth]{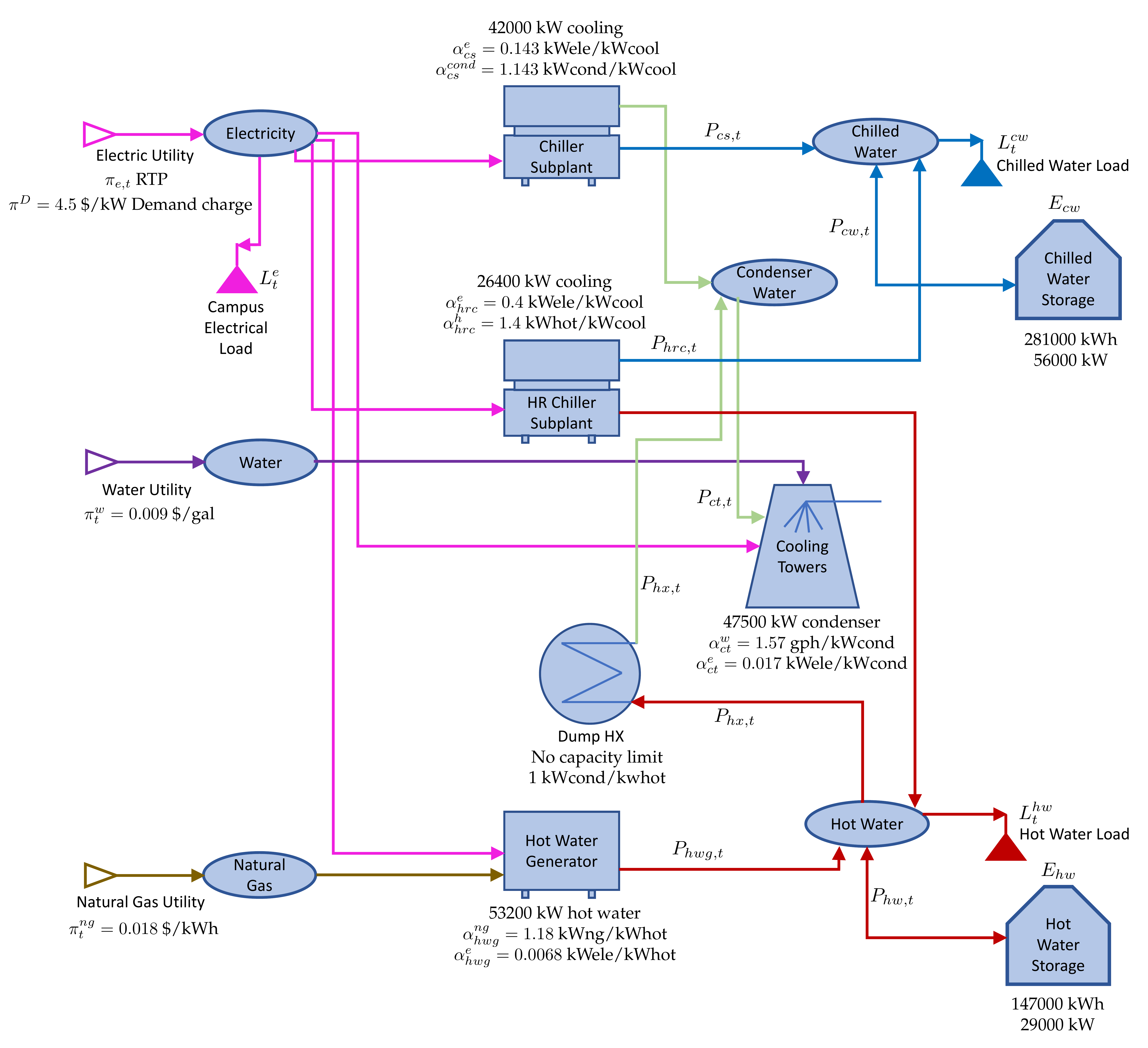}
\caption{Schematic energy flow diagram of central HVAC plant (reproduced with permission) \cite{kumar2020stochastic}.}
\label{fig:schematic}
\end{figure}

Figure \ref{fig:schematic} shows the energy flows between all units of the HVAC plant and interactions with loads and utilities. As in Figure \ref{fig:schematic}, the amount of electricity, water, and natural gas consumed by the units depends on their operating loads. The chiller and HR chiller subplants use $\alpha^{e}_{cs}$ and $\alpha^{e}_{hrc}$ kW of electricity for the production of 1 kW of chilled water, respectively; the hot water generator requires $\alpha^{e}_{hwg}$ kW of electricity and $\alpha^{ng}_{hwg}$ kW of natural gas for the production of 1 kW of hot water; and the cooling towers require $\alpha^{e}_{ct}$ kW of electricity and $\alpha^{w}_{ct}$ utility water for 1 kW of condenser water input. For the chilled water load of the campus ($L^{cw}_{t}$), chilled water is produced by the chiller ($P_{cs,t}$), the HR chiller subplants ($P_{hrc,t}$), and the discharge from chilled water storage ($P_{cw,t}$). For the hot water load of the campus ($L^{hw}_{t}$), hot water is produced by the HR chiller subplant ($\alpha^{h}_{hrc}P_{hrc,t}$), the hot water generator ($P_{hwg,t}$), and the discharge from the hot water storage ($P_{hw,t}$). The excess hot water ($P_{hx,t}$) in the system is recycled by cooling it and producing condenser water in the dump HX, and the cooling towers use the evaporative cooling to reduce the temperature of this condenser water along with the condenser water produced by the chiller and the HR chiller subplants (total $P_{ct,t}$ condenser water). 

In the MPC formulation, the operating loads of all units of the HVAC plant are the manipulated variables, while the states include the state of charge (SOC) of the chilled water and hot water storage tanks (TES) and carryover quantities (e.g., peak electrical demand, unmet or overmet production of chilled/hot water). Multiple time-varying disturbances are present in this system; these include the campus electrical load ($L^{e}_{t}$), chilled water load ($L^{cw}_{t}$), hot water load ($L^{hw}_{t}$), and electricity prices ($\pi^{e}_{t}$). The MPC uses forecasts for these disturbances over a prediction horizon $\mathcal{T}$ to determine the control action for the next immediate hour. The horizon is shifted by one hour to update disturbance forecasts and to obtain the next control action. This procedure is repeated for an entire year to obtain the closed-loop policy and associated cost.  The optimization problem solved at each time $t$ is: 

{\setlength{\mathindent}{0pt}
\begin{subequations} \label{Eq:detmpc}
\begin{align}
\min \; & \sum\limits_{k \in \mathcal{T}}  \sum\limits_{j=\{e,w,ng\}}\hat{\pi}^j_k r^{j}_{k} + \frac{\pi^D}{\sigma_t} R_{t+1} \notag \\ & + \sum\limits_{k \in \mathcal{T}}\sum\limits_{j \in \{cw,hw\}} \rho_{j}(ul_{j,k}+ol_{j,k}). \label{eq:objective}\\
\textrm{s.t. }\; & r^{e}_{k} = \sum_{j \in \{cs,hrc,hwg,ct\}} \alpha^{e}_{j} P_{j,k} + \hat{L}^{e}_{k}, \; k \in \mathcal{T}\label{eq:elec_load}\\
& r^{j}_{k} = \alpha^{j}_{i_j} P_{i_j,k}, \; j \in \{w,ng\}, k \in \mathcal{T}, \; i_{w}=ct, i_{ng}=hwg  \label{eq:water_ng_load} \\
& P_{ct,k} = \alpha^{cond}_{cs} P_{cs,k}+P_{hx,k}, \; k \in \mathcal{T} \label{eq:dumphx} \\
& P_{cs,k}+P_{hrc,k}+P_{cw,k}+S^{un}_{cw,k}-S^{ov}_{cw,k} = \hat{L}^{cw}_{k}, \; k \in \mathcal{T} \label{eq:cw_load} \\
& \alpha^{h}_{hrc} P_{hrc,k}+P_{hwg,k}-P_{hx,k}+P_{hw,k} \notag \\ &+S^{un}_{hw,k}-S^{ov}_{hw,k}  = \hat{L}^{hw}_{k}, \; k \in \mathcal{T} \label{eq:hw_load} \\
& E_{j,k+1} = E_{j,k} - P_{j,k}, \; j \in \{cw,hw\}, k \in \mathcal{T} \label{eq:edynamics} \\
& ul_{j,k+1} = ul_{j,k} - S^{m}_{j,k}, \; m \in \{un, ov\}, j \in \{cw, hw\}, \notag \\ & \phantom{ul_{j,k+1} = ul_{j,k} - S^{m}_{j,k}, \; m \in \{un, ov\}, } k \in \mathcal{T}\label{eq:uldynamics} \\
& ol_{j,k+1} = ol_{j,k} - S^{m}_{j,k}, \;  m \in \{un, ov\}, j \in \{cw, hw\}, \notag \\ & \phantom{ol_{j,k+1} = ol_{j,k} - S^{m}_{j,k}, \;  m \in \{un, ov\}, } k \in \mathcal{T} \label{eq:oldynamics} \\
& R_{t+1} \geq r^{e}_{k}  \label{eq:peakdemand} \\
& R_{t+1} \geq R_{t}  \label{eq:peakdemand_carry} \\
& \underline{E}_{j,k} \leq E_{j,k} \leq \overline{E}_{j,k}, \; j \in \{cw,hw\}, k \in \mathcal{T} \label{eq:ebound}\\
& \underline{P}_j \leq P_{j,k} \leq \overline{P}_j, \; j \in \{cs,hrc,hwg,ct,hx,cw,hw\}, \notag \\ & \phantom{\underline{P}_j \leq P_{j,k} \leq \overline{P}_j, \; } k \in \mathcal{T} \label{eq:p1bound}\\
& S^{m}_{j,k} \geq 0, \;  m \in \{un, ov\}, j \in \{cw, hw\},  k \in \mathcal{T} \label{eq:Sbound}\\
& ul_{j,k} \geq 0, \; j \in \{cw, hw\}, k \in \mathcal{T}   \label{eq:ulbound}\\
& ol_{j,k} \geq 0, \; j \in \{cw, hw\}, k \in \mathcal{T}  \label{eq:olbound}
\end{align}
\end{subequations}}
Here, the residual demands of electricity, water, and natural gas that need to be purchased from the markets are given by the constraints \eqref{eq:elec_load}-\eqref{eq:water_ng_load}. Constraints \eqref{eq:dumphx}-\eqref{eq:hw_load} are the energy balance equations for the condenser water. The sufficient chilled and hot water production is maintained by imposing constraints \eqref{eq:cw_load} and \eqref{eq:hw_load} (with some slack variables for under-production or over-production for feasibility). The state variables $ul_{j,k}$ and $ol_{j,k}$, $j \in \{cw, hw\}$ carry over the under-production or over-production of chilled and hot water in constraints \eqref{eq:uldynamics} and \eqref{eq:oldynamics} and these state variables are penalized in the objective function. The dynamics of SOC for chilled and hot water TES are given by constraints \eqref{eq:edynamics}. Constraints \eqref{eq:peakdemand} compute the peak demand over the horizon and constraint \eqref{eq:peakdemand_carry} carries over the peak demand to the next time step in the closed-loop. 

The actual realizations of the loads (disturbances) might induce constraint violations when they deviate from the forecasts. To account for such violations,  bounds on the chilled and hot water TES in \eqref{eq:ebound} are modified to include a buffer capacity (the back-off term), $\beta_j \in [0,0.5], j \in \{cw,hw\}$. In closed-loop, the bounds on $E_{j,k}$ for $j \in \{cw,hw\}$ in constraints \eqref{eq:ebound} are updated as: \\
If $\beta_j \overline{E}_j \leq E_{j,t+1} \leq (1-\beta_j) \overline{E}_j$, set $\underline{E}_{j,t+1} = \beta_j \overline{E}_j$, $\overline{E}_{j,t+1} = (1-\beta_j) \overline{E}_j$.\\
If $(1-\beta_j) \overline{E}_j \leq E_{j,t+1} \leq \overline{E}_j$, set  $\underline{E}_{j,t+1} = \beta_j \overline{E}_j$, $\overline{E}_{j,t+1} = E_{j,t+1}$. \\
If $0 \leq E_{j,t+1} \leq \beta_j \overline{E}_j$, set  $\underline{E}_{j,t+1} = E_{j,t+1} $, $\overline{E}_{j,t+1} = (1-\beta_j) \overline{E}_j$. \\
If $E_{j,t+1} \geq \overline{E}_j$,set $E_{j,t+1} =\overline{E}_j$, $\underline{E}_{j,t+1} = \beta_j \overline{E}_j $, $\overline{E}_{j,t+1} = \overline{E}_j$, and update $ol_{j,k+1}=ol_{j,k+1}+(E_{j,t+1} - \overline{E}_j)$. \\
If $E_{j,t+1} \leq 0$, set $E_{j,t+1} =0$, $\underline{E}_{j,t+1} = 0$, $\overline{E}_{j,t+1} = (1-\beta_j)\overline{E}_j$, and update $ul_{j,k+1}=ul_{j,k+1}-E_{j,t+1}$.\\
The above updates to the storage bounds ensure that, if the storage at $t+1$ overflows or dries up when implementing the MPC action, the storage is set to the maximum or minimum capacity, respectively;  otherwise, the fractional buffer capacity is implemented. These corrections result in lost economic performance and inefficient use of storage. We perform closed-loop MPC simulations for the central HVAC plant with the formulation described above and develop the BO framework for tuning the back-off terms for the chilled water and hot water TES. 
\\

The back-off term $\mathbf{\beta}=[\beta_{cw},\beta_{hw}]^{T}$ is introduced  to reserve a $\beta$ fraction of the maximum capacity as the buffer to account for the unpredictable disturbance uncertainty. Appropriate determination of the $\beta$ value is critical to maximize closed-loop performance. If $\beta$ is too large, this will induce an overly conservative strategy that prevents storage tanks from being fully utilized to reduce economic cost. On the other hand, if $\beta$ is too small, the number of constraint violations may increase dramatically since there is not enough buffer to provide a safeguard against unforeseen disturbances, leading to an economic penalty. The back-off parameter $\beta$ affects closed-loop MPC performance in a non-intuitive way. For this study, our tuning objective is the annual closed-loop cost  (denoted as $f(\cdot)$), which is a function of the back-off terms 
$\xi=\{\beta_{cw}$, $\beta_{hw}\}$. A year-long closed-loop simulation for the HVAC plant has to be performed to evaluate $f(\xi)$.

\begin{figure}
	\centering
	\includegraphics[width=0.8\linewidth]{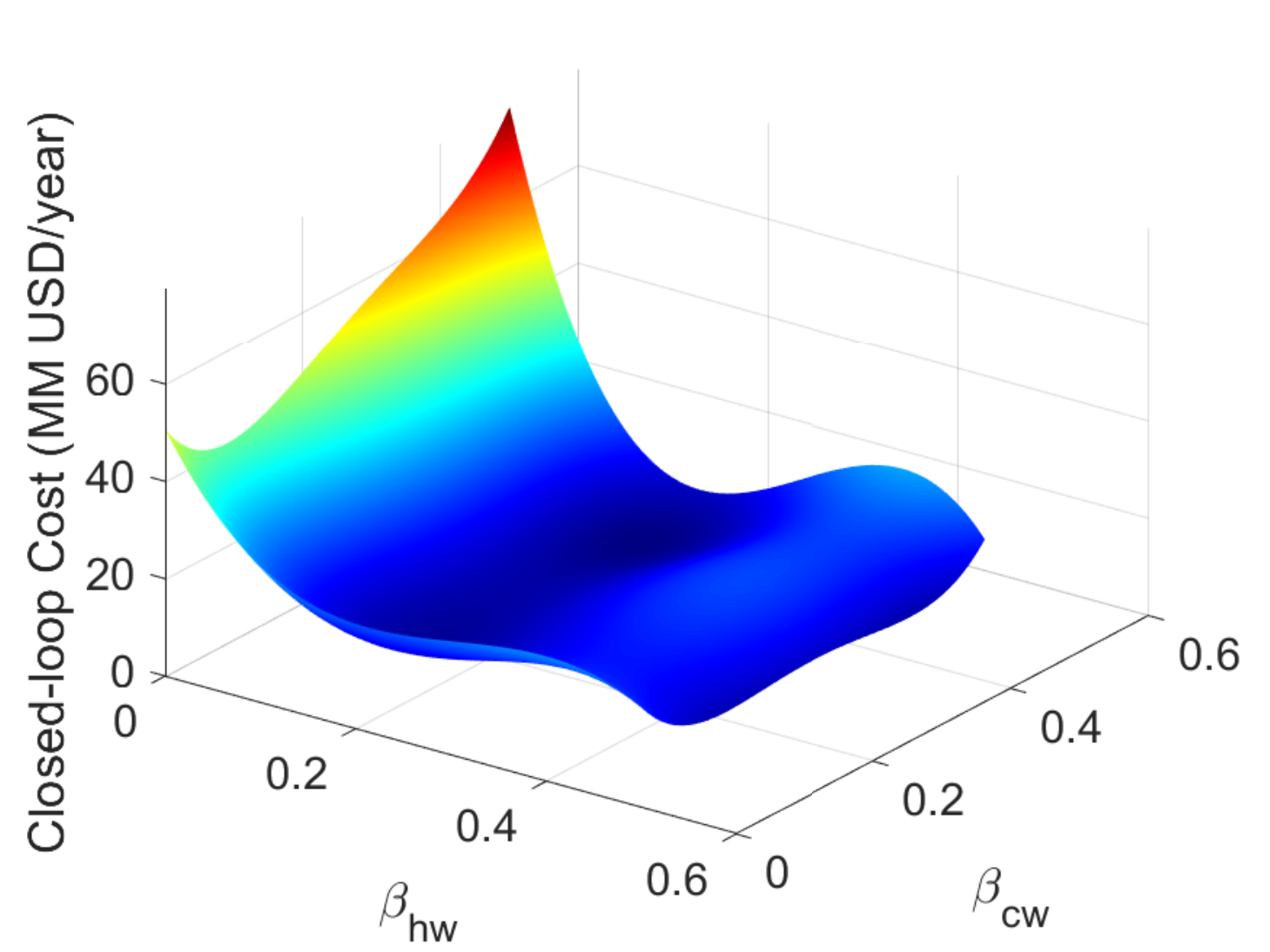}
	\caption{Year-long closed-loop cost for HVAC plant under different $\beta_{cw}$ and $\beta_{hw}$ values (interpolated surface based on values of sampled grids).}
	\label{fig: Total_cost_3d}
\end{figure}

The prediction horizon of MPC is chosen to be 168 hours (1 week) to reflect the weekly periodicity of loads and electricity prices. The optimization problem solved at each hour is a linear program with 168,450 variables and 143,750 constraints \cite{kumar2020stochastic}. The problems were implemented in Julia 0.6.4 and were solved with Gurobi 8.1 on a computing server with 188 GB RAM, 32-core Intel Xeon 2.30 GHz CPU. On average, each MPC problem requires about one second to solve but simulating closed-loop behavior over an entire year requires about 2 hours of wall-clock time (each year-long simulation requires solving more than 8,700 optimization problems). Given the complexity of the underlying tuning problem, it is apparent that manual search or grid search method is not applicable due to the possibly large number of trials and the resultant enormous time consumed.  

Figure \ref{fig: Total_cost_3d} shows the closed-loop cost with different combinations of back-off values for a given disturbance realization. To generate this surface, we conducted 81 simulations (obtained by using a coarse grid discretization with 9 points for each back-off term). One can see that the surface is non-convex with a couple of local minima (the global minimum is near $\beta_{cw}=0.4$, $\beta_{hw}=0.2$).  Note also that the closed-loop cost is highly sensitive to the back-off terms. The large costs also illustrate that operating HVAC facilities is quite expensive and thus cutting down costs is essential.  {\em To simplify our computational analysis, we used this surface as the ``real" objective function (i.e., we do not conduct closed-loop simulations during the BO search). As such, the objective surface might not be representative of actual costs of the system (it is only used to illustrate the algorithmic performance of BO).} We used the back-off term values $\beta_{cw}=\beta_{hw}=0.1$ reported in \cite{kumar2020stochastic} as a baseline. Note that the selected parameters above may not be optimal for our case study due to a different disturbance realization used in this work. 

\begin{figure}
	\centering
	\includegraphics[width=0.8\linewidth]{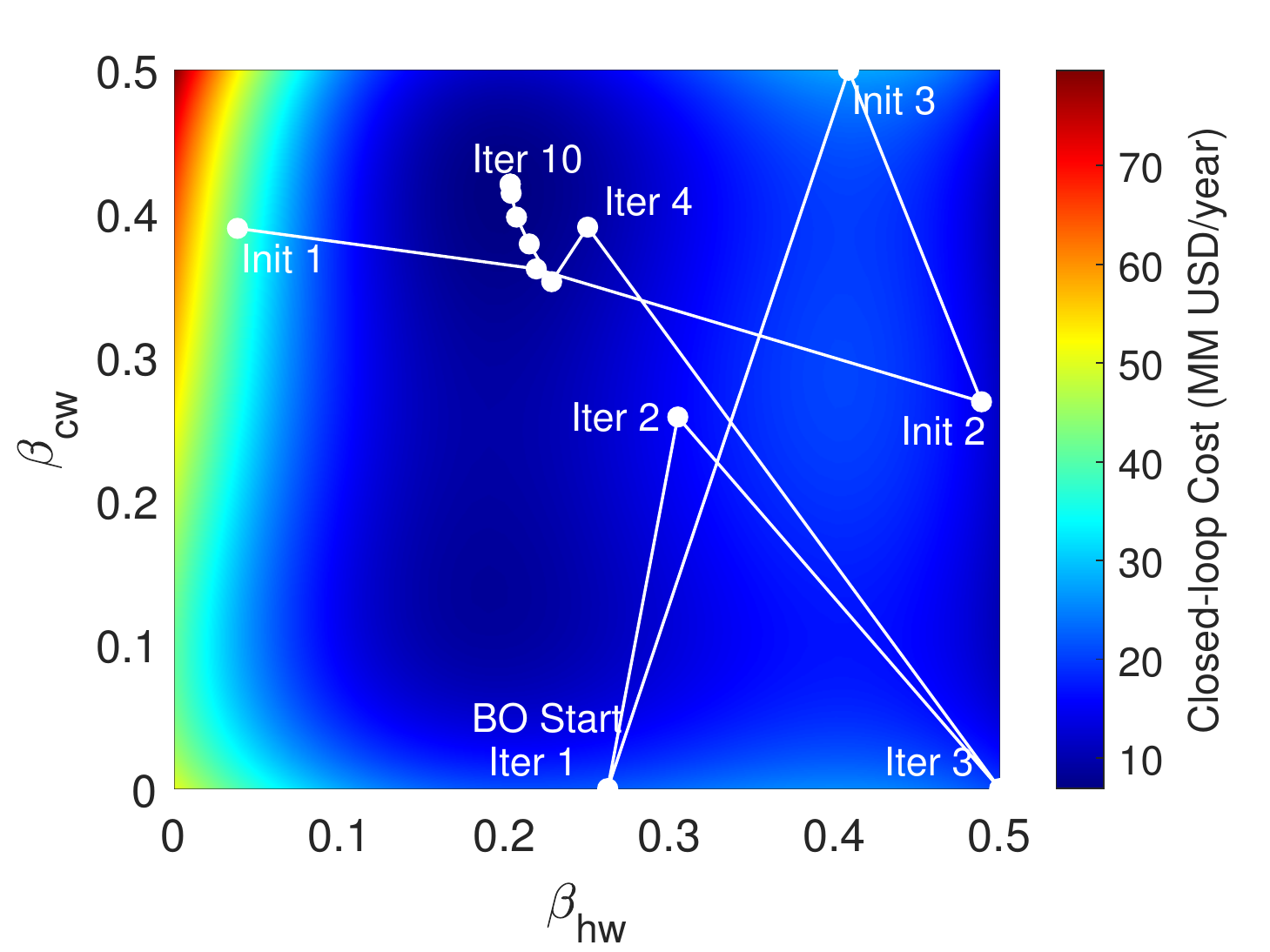}
	\caption{The sequence of optimal sampling locations in the iterations of Bayesian optimization.}
	\label{fig: Iterations}
\end{figure}

\begin{figure}[h]
	\centering
	\includegraphics[width=0.8\linewidth]{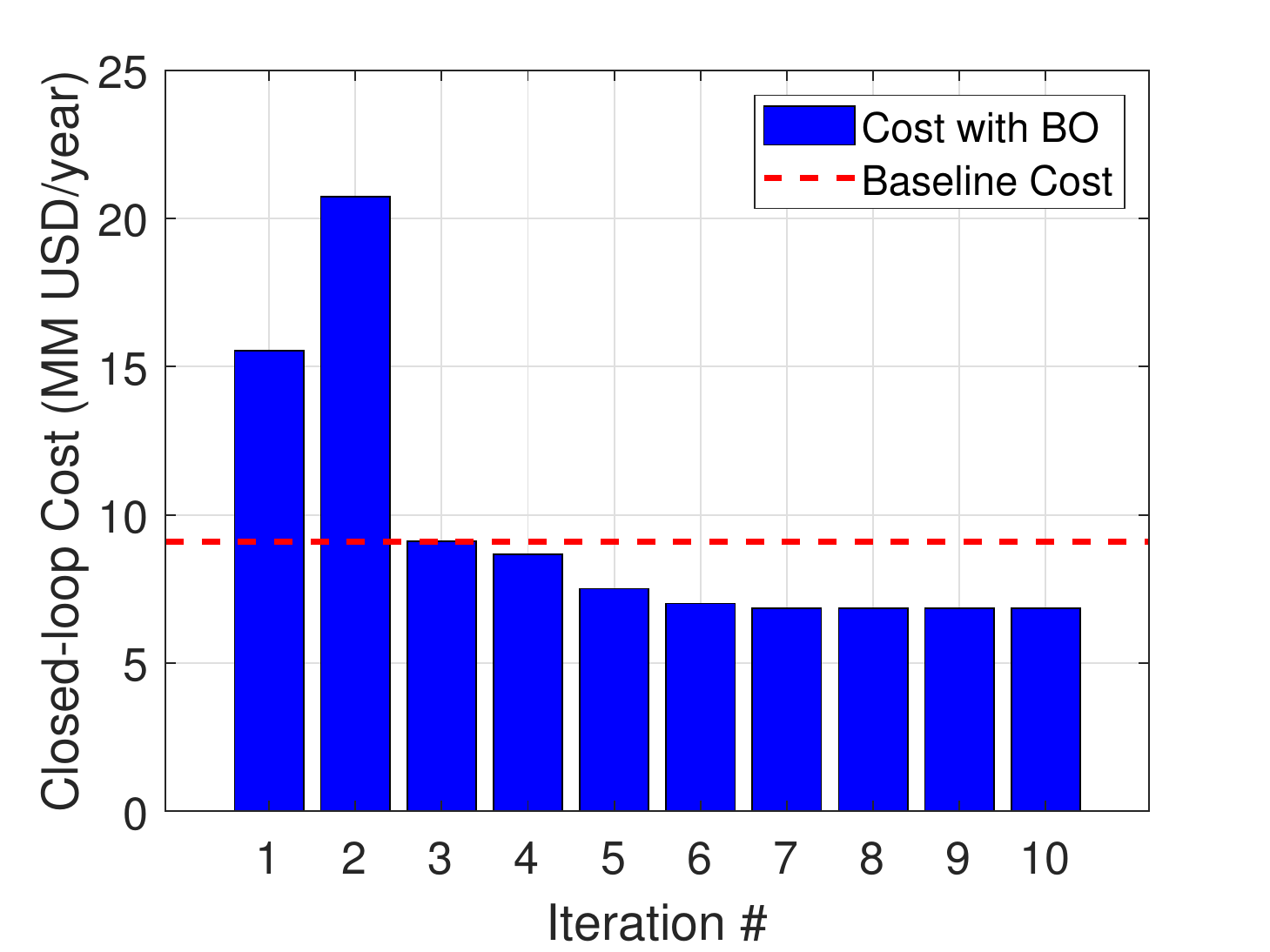}
	\caption{The closed-loop cost over Bayesian optimization iterations.}
	\label{fig: Total_cost}
\end{figure}

\begin{figure*}[tb]
	\centering
	\includegraphics[width=0.7\linewidth]{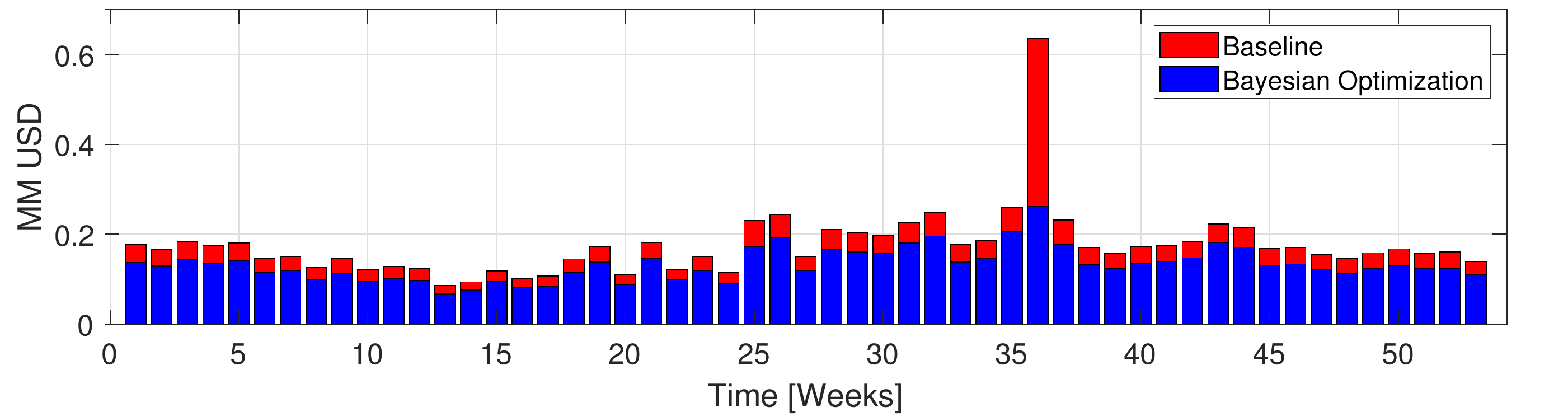}
	\caption{Week closed-loop costs for baseline and tuned back-off terms with Bayesian optimization.}
	\label{fig: Weekly_Cost}
\end{figure*}

We used BO to determine the optimal back-off terms for the MPC controller. The algorithm is programmed in Python 3.8.3 using the pyGPGO package\footnote{Available at https://github.com/josejimenezluna/pyGPGO} with a ``matern52'' kernel function ($l=1, \nu=5/2, \sigma^2=1\mathrm{e}{-6}$) for GP model and UCB AF ($\kappa=2.6$). The optimizer for finding the maximum of the AF is chosen as the limited-memory BFGS (L-BFGS) algorithm. Starting with $n=3$ initial points, the sequence of optimal sampling points delivered by BO is shown in Figure \ref{fig: Iterations}.  One can see that, after three iterations of exploration, BO  starts to converge to a neighborhood of the global minimum and convergence is achieved in only 10 iterations. The total number of closed-loop simulations evaluations, including the initial points, is only 13 for this example, significantly reducing the amount of  computation involved (compared to the coarse grid search used to generate Figure \ref{fig: Total_cost_3d}). Figure \ref{fig: Total_cost} shows the closed-loop cost of the HVAC plant at each BO iteration. One can see a monotonically decreasing trend after initial exploration (which increases cost). This highlights that high performance gains that can be achieved with MPC tuning.  Figure \ref{fig: Weekly_Cost} shows the weekly operation costs corresponding to the optimal BO parameters  and with the baseline parameters. We can see that the tuned parameters achieve weekly costs that are consistently lower than those of the baseline (largest savings occur in late August). 

\begin{figure*}[tb]
	\centering
	\includegraphics[width=0.7\linewidth]{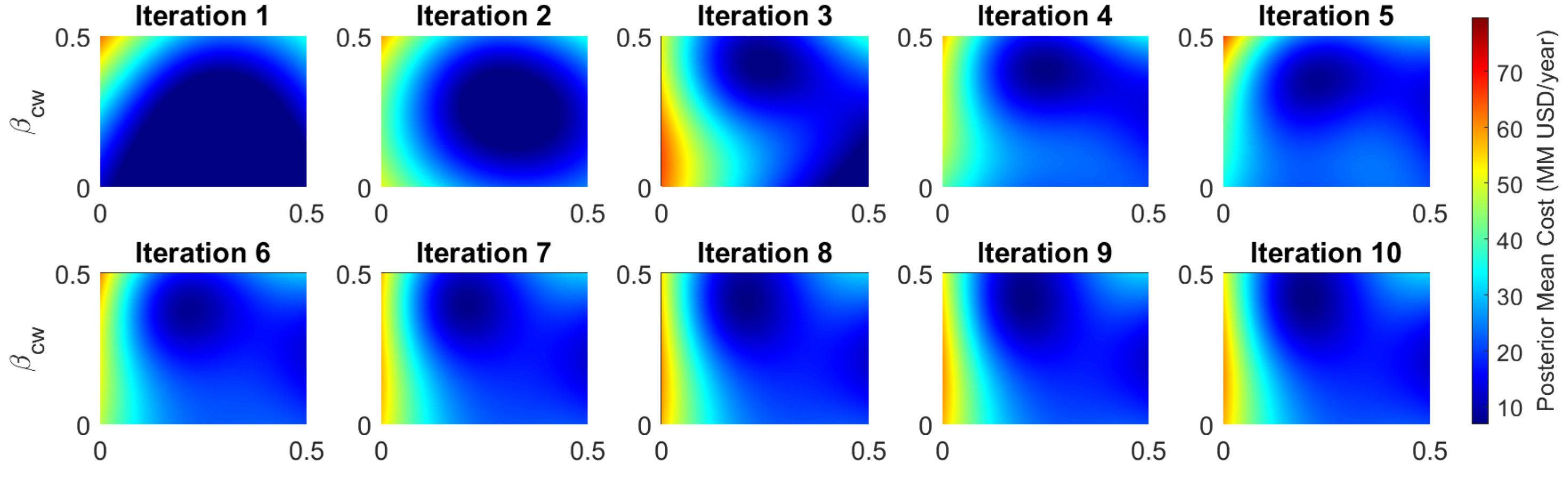}
	\includegraphics[width=0.7\linewidth]{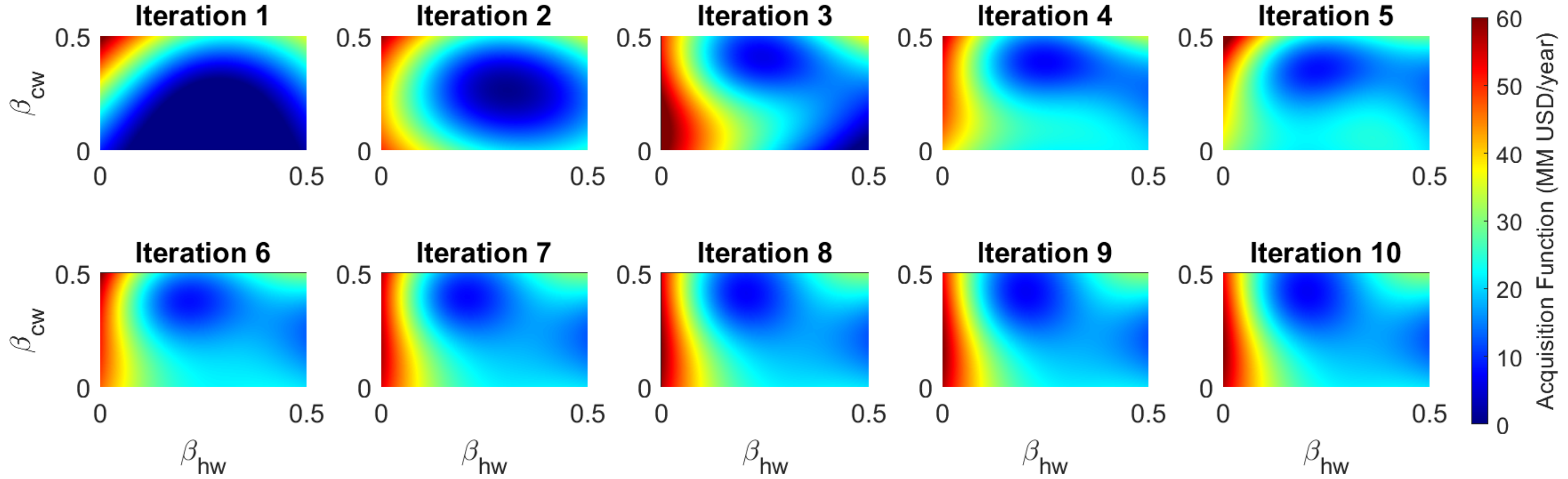}
	\caption{Top two rows: Posterior mean of GP model in each iteration. Bottom two rows: Acquisition function in each iteration.}
	\label{fig: Posterior_Acquisition}
\end{figure*}

The posterior mean of the GP model in each iteration and the corresponding AF values are shown in Figure \ref{fig: Posterior_Acquisition}. The top two rows of Figure \ref{fig: Posterior_Acquisition} show that, after a few iterations, the posterior mean surface becomes consistent, indicating convergence of the algorithm.  The bottom two rows present the AF in each iteration. It is interesting to see that, after several iterations, the minimum of the AF stays near the global solution and the neighborhood of the other local minimum does not present large AF values. It is thus anticipated that the BO iteration is unlikely to jump to the other local minimum. 

In summary, our results indicate that BO can achieve significant reductions in cost by identifying optimal back-off terms for chilled and hot water tanks in the HVAC plant. The convergence of the algorithm can be achieved using a few year-long closed-loop simulations.

\section{Conclusions and Future Work}
We presented a BO framework for tuning MPC controllers. The tuning objectives are treated as a black-box function of the controller parameters. This work is motivated by the observation that evaluating closed-loop performance can be computationally expensive and thus manual or grid search approaches are time-consuming. BO is used to efficiently solve this complex MPC tuning problem; specifically, we studied the optimization of the back-off terms for the thermal energy storage of an HVAC plant. Our results show that BO can effectively find back-off terms by performing 13 closed-loop simulations and this can save total costs. As part of future work, we are interested in exploring performance with a larger set of tuning parameters that capture different types of behavior and different types of  functions to accelerate the search.  

\section*{ACKNOWLEDGMENT}

We acknowledge the support of the members of the Texas-Wisconsin-California control consortium.






\end{document}